\documentclass[12pt,preprint]{aastex}
\usepackage{natbib}
\usepackage{graphicx}
\newcommand{\msun} {$M_{\odot}$}
\newcommand{\me}{$M_{\oplus}$}

\newcommand{\Te} {$T_{\rm eff}$}
\newcommand{\logg} {$\log g$}

\begin{document}

\slugcomment{\bf Accepted for publication in ApJ}

\title{LIMB-DARKENING COEFFICIENTS FOR ECLIPSING WHITE DWARFS}

\author{A. Gianninas$^{1}$, B. D. Strickland$^{1}$, Mukremin Kilic$^{1}$
  and P. Bergeron$^{2}$}

\affil{$^{1}$Homer L. Dodge Department of Physics and Astronomy,
  University of Oklahoma, 440~W.~Brooks~St., Norman, OK 73019,
  USA}\email{alexg@nhn.ou.edu, benstrickland@ou.edu, kilic@ou.edu}
\affil{$^{2}$D\'epartement de Physique, Universit\'e de Montr\'eal,
  C.P.~6128, Succ.~Centre-Ville, Montr\'eal, Qu\'ebec H3C 3J7,
  Canada}\email{bergeron@astro.umontreal.ca}

\begin{abstract}

  We present extensive calculations of linear and non-linear
  limb-darkening coefficients as well as complete intensity profiles
  appropriate for modeling the light-curves of eclipsing white
  dwarfs. We compute limb-darkening coefficients in the
  Johnson-Kron-Cousins {\it UBVRI} photometric system as well as the
  Large Synoptic Survey Telescope (LSST) {\it ugrizy} system using the
  most up-to-date model atmospheres available. In all, we provide the
  coefficients for seven different limb-darkening laws. We describe
  the variations of these coefficients as a function of the
  atmospheric parameters, including the effects of convection at low
  effective temperatures. Finally, we discuss the importance of having
  readily available limb-darkening coefficients in the context of
  present and future photometric surveys like the LSST, Palomar
  Transient Factory, and the Panoramic Survey Telescope and Rapid
  Response System (Pan-STARRS).  The LSST, for example, may find
  $\sim10^5$ eclipsing white dwarfs.  The limb-darkening calculations
  presented here will be an essential part of the detailed analysis of
  all of these systems.

\end{abstract}


\section{MOTIVATION}

The study of binary systems is of great importance in astrophysics
since it allows us to directly measure stellar parameters based solely
on our understanding of the dynamics of the system. Eclipsing systems
are even more valuable as they provide model-independent estimates of
the properties of both objects in the system, such as the stellar (or
planetary) masses and radii. Of particular interest here is the
discovery in the past several years of an increasing number of
eclipsing binary systems including white dwarf stars.

First, the number of known white dwarfs with main-sequence M dwarf
companions has exploded \citep[e.g.,][]{silvestri06,rm12} with the
continuing discoveries based on the Sloan Digital Sky Survey (SDSS)
data. In addition, new transient surveys like the Catalina Real-time
Transient Survey \citep{drake09} and the PTF/M-dwarf transiting planet
survey \citep{law12} are finding more and more eclipsing white dwarf
plus M dwarf systems. Indeed, there are now 49 eclipsing DA+dM binary
systems known
\citep{nebot09,parsons12a,parsons12b,parsons12c,pyrzas09,pyrzas12}.
This has led to some of the most precise determinations of
model-independent masses and radii for white dwarfs and M dwarfs. Such
results are of prime importance as they allow for a confirmation of
the mass--radius relation for white dwarfs. With a measurement of the
surface gravity from optical spectroscopy, the mass--radius relation
provides mass estimates for isolated white dwarfs.

Double-degenerate binary systems composed of two white dwarfs stars
are even more important, as they provide independent mass--radius
measurements for both white dwarfs in the system. \citet{steinfadt10}
reported the discovery of the first ever eclipsing double white dwarf
system, NLTT~11748. This system contains an $\approx$0.18 \msun\ and a
0.76 \msun\ companion in a 5.6 hr orbit
\citep{steinfadt10,kawka10,kilic10}. \citet{parsons11} identified
CSS~41177 as another eclipsing double white dwarf system containing
two 0.27--0.28 \msun\ white dwarfs in a 2.8 hr orbit. More recently,
the extremely low mass (ELM) survey
\citep{brown_ELM1,kilic_ELM2,brown_ELM3,kilic_ELM4}, a targeted search
for ELM ($M \sim$ 0.25 \msun) white dwarfs has uncovered more than 50
short period ($P < 1$~day) white dwarf binaries
\citep{brown_ELM5}. This eventually led to the discovery of J0651, a
765 s orbital period eclipsing detached double white dwarf binary
containing a 0.26 \msun\ white dwarf and a 0.50 \msun\ companion
\citep{brown11,hermes12b}.  Finally, \citet{vennes11} identified
another ELM white dwarf, GALEX J171708.5+675712, as an eclipsing
detached double white dwarf system with an orbital period of 5.9
hr. So far, all of the known, eclipsing, detached, double white dwarf
systems involve low-mass white dwarfs. This is most likely because it
is easier to identify eclipsing white dwarfs in short period binaries,
which form as a result of one or two phases of common envelope
evolution that leads to the formation of low-mass white dwarfs.

Due to their small sizes ($R\sim 0.01 R_{\odot}$), the eclipse
probability is relatively small for white dwarfs. \citet{dstef10}
estimate that the probability of eclipse by a planet at 1~AU from an
average size white dwarf is only 0.0002. Hence, the discovery of a
large number of eclipsing white dwarfs requires wide-field transient
surveys that will image thousands of white dwarfs. In the future, we
expect projects like the Large Synoptic Survey Telescope
\citep[LSST;][]{lsst08}, Palomar Transient Factory
\citep[PTF;][]{law09,rau09}, and Panoramic Survey Telescope and Rapid
Response System \citep[Pan-STARRS;][]{tonry12} to uncover thousands of
new eclipsing binary white dwarfs systems, including double white
dwarf and white dwarf + M dwarf systems. An even more enticing
prospect is the possibility that these surveys might uncover
transiting planets around white dwarfs.

Modern instrumentation affords sufficient precision that we need to
take into account all the necessary physics in order to properly model
the observed light-curves of eclipsing systems. Ellipsoidal variations
in tidally distorted systems \citep{kilic11,brown11,hermes12a} and the
relativistic beaming effect \citep{loeb03,zucker07,shporer10} can
affect the precise shapes of the light-curves. The exact shape of the
eclipses and the amplitude of the ellipsoidal variations depend on the
effects of limb-darkening, which is important for precise measurements
of masses and radii for eclipsing white dwarfs and their stellar and
planetary companions.

There exist many published tables of limb-darkening coefficients for
main-sequence stars based on various models and covering a variety of
atmospheric parameters, limb-darkening laws, and photometric
systems. Since there was never a pressing need, analogous tables
appropriate for white dwarf stars have never been published. With the
increasing number of known white dwarf eclipsing binaries, and the
expectation of many more such discoveries in the coming years, it is
an opportune time to provide the community with a useful resource
through the publication of linear and non-linear limb-darkening
coefficients for white dwarfs.

In Section 2, we present the model atmospheres which we use as a basis
for the computation of the limb-darkening coefficients. In Section 3,
we discuss seven widely-used limb-darkening laws. Next, in Section 4
we present a few representative intensity profile fits and present our
complete tables of limb-darkening coefficients. In Section 5 we
explore the behavior of the limb-darkening coefficients as a function
of \Te\ and \logg\ and we make suggestions as to which limb-darkening
laws are best in different \Te\ and \logg\ regimes.  In Sections 6 and
7, we discuss the transient survey yields for eclipsing white dwarfs
with stellar and planetary mass companions.  Finally, in Section 8 we
summarize our work.

\section{MODEL ATMOSPHERES}

Since over 80\% of white dwarfs have hydrogen-rich atmospheres we
restrict our calculations to hydrogen-rich model atmospheres. We note,
however, that all the work described below can just as easily be
applied to any other type of white dwarf atmosphere (helium, carbon),
should the need arise. Our models are described at length in
\citet{TB09} and references therein, but for the sake of completeness
we briefly describe them here. Our models assume a plane-parallel
geometry and are computed in local thermodynamic equilibrium
(LTE). The assumption of LTE is warranted as our models are restricted
to \Te\ $<$ 30,000~K. In particular, our models include the new Stark
broadening profiles of \citet{TB09} which include the occupation
probability formalism of \citet{hm88} directly in the line profile
calculation. Furthermore, these models employ the ML2/$\alpha$ = 0.8
version of the mixing-length theory, as prescribed by
\citet{tremblay10}, for models where convective energy transport
becomes important.

The main difference between the grid of models presented in
\citet{TB09} and \citet{tremblay10} and the one used here is the
extent of our model grid in \Te\ and \logg. The grid used here spans
effective temperatures from \Te\ = 4000~K to 30,000~K in steps varying
from 250~K to 5000~K, and surface gravities ranging from \logg\ = 5.0
to 9.5 in steps of 0.25 dex. The new grid extends to much lower
surface gravities because we initially computed it for the analysis of
the white dwarfs from the ELM survey. Since the ELM survey itself has
produced one eclipsing white dwarf binary, our initial calculations
were done for that system only \citep[J0651][]{hermes12b}. Here we
present limb-darkening coefficients for our entire grid of models.
The above range of \Te\ encompasses the vast majority of white dwarfs
that the LSST and other transient surveys will observe.  For example,
the Besan\c con Galaxy model \citep{robin03} predicts 56 white dwarfs
per square degree at a Galactic latitude of 59$\arcdeg$ down to $g=24$
mag. Only one of these 56 stars is outside the lower boundary of our
temperature grid. Our calculations can be extended to hotter and
cooler temperatures, if necessary.

\section{LIMB-DARKENING LAWS}

In order to provide the most comprehensive tabulation of linear and
non-linear limb-darkening coefficients that encompasses as many needs
and uses as possible, we consider a total of seven different
limb-darkening laws. In all cases, $I(1)$ represents the specific
intensity at the center of the stellar disk and $a$, $b$, $c$, $d$,
$f$, $g$, $h$, $j$, $k$, $l$, $m$, $n$, $p$, $q$, $r$, $s$ are the
limb-darkening coefficients for different laws. The most basic law is
the linear limb-darkening law that can be traced all the way back to
\citet{r1912} and takes the form

\begin{equation}
\frac{I(\mu)}{I(1)} = 1 - a(1-\mu),
\end{equation}

\noindent where $\mu = \cos{\theta}$ and $\theta$ is the angle between
the line of sight and the direction of the emergent flux. However,
limb-darkening is usually not a simple linear function of
$\mu$. Nonlinear laws tend to reproduce the intensity profile of the
atmosphere better than a linear limb-darkening law. Three of the most
common nonlinear limb-darkening laws are the quadratic
\citep{kopal50}, square-root \citep{diaz92}, and logarithmic
\citep{ks70} laws which are defined in Equations(2) through (4),
respectively,

\begin{equation}
\frac{I(\mu)}{I(1)} = 1 - b(1-\mu) - c(1-\mu)^{2}
\end{equation}

\begin{equation}
\frac{I(\mu)}{I(1)} = 1 - d(1-\mu) - f(1-\sqrt{\mu})
\end{equation}

\begin{equation}
\frac{I(\mu)}{I(1)} = 1 - g(1-\mu) - h(\mu \ln{\mu}).
\end{equation}

More recently, several investigators have defined new nonlinear
limb-darkening laws with more parameters. For example,
\citet{claret03} introduced an exponential limb-darkening law, as
given in Equation (5):

\begin{equation}
\frac{I(\mu)}{I(1)} = 1 - j(1-\mu) - \frac{k}{(1-e^{\mu})}.
\end{equation}

\noindent Furthermore, \citet{sing09} defined a three-parameter law
given in Equation (6):

\begin{equation}
\frac{I(\mu)}{I(1)} = 1 - l(1-\mu) - m(1-\mu^{\frac{3}{2}}) - n(1-\mu^{2}).
\end{equation}

\noindent The above 3-parameter law is based on the 4-parameter law,
defined in Equation (7), from \citet{claret00} with the
$\mu^{\frac{1}{2}}$ term omitted:

\begin{equation}
\frac{I(\mu)}{I(1)} = 1 - p(1-\mu^{\frac{1}{2}}) - q(1-\mu) - r(1-\mu^{\frac{3}{2}}) -s(1-\mu^{2}).
\end{equation}

\section{LIMB-DARKENING COEFFICIENTS}

The first step in determining the limb-darkening coefficients is the
computation of intensity profiles based on our model atmospheres. To
calculate the specific intensities at the surface $I(\nu, \mu,
\tau_{\nu} = 0)$, we solve the radiative transfer equation for various
values of $\mu$.  We compute specific intensities for 101 evenly
spaced values between $\mu$ = 0 (the edge of the disk) and $\mu$ = 1
(the center of the disk).  These intensity profiles are computed for
wavelengths from $\lambda$ = 2500~\AA\ to 95000~\AA.

To compute the limb-darkening coefficients for specific filters, the
next step is to integrate $I(\nu,\mu)$, or alternatively
$I(\lambda,\mu)$, over these bandpasses using the filter transmission
functions. In this paper, we present our calculations for two widely
used photometric systems; the broadband Johnson--Kron--Cousins $UBVRI$
system and the $ugrizy$ system to be used by the LSST.  For the
$UBVRI$ filters, we use the filter transmission functions from
\citet{landolt92a,landolt92b}. For the $ugrizy$ system, we adopt the
LSST $ugrizy$ filters, which are based on the SDSS $ugriz$ filters and
include the additional filter $y$ at 1~$\mu m$. We use the LSST filter
transmission curves from the LSST Survey Science Group Web
site.\footnote{ssg.astro.washington.edu/elsst/elsst.html} We integrate
over the filter bandpasses according to the following equation:

\begin{equation}
I(\mu)=\frac{\int I(\lambda,\mu)S(\lambda)\lambda d\lambda}{\int S(\lambda)\lambda d\lambda},
\end{equation}

\noindent where $S(\lambda)$ is the filter transmission function, and
$I(\lambda,\mu)$ is the monochromatic specific intensity. We perform
the integration over the filter bandpass in wavelength (in
angstroms). We use the integrated specific intensity $I(\mu)$ to
determine the limb-darkening coefficients for every \Te\ and \logg\ in
our model grid. There are a total of 11 filters ($UBVRI$ and $ugrizy$)
and 16 coefficients for the seven limb-darkening laws that we have
chosen.

We constrain the limb-darkening coefficients using the nonlinear
least-squares method of Levenberg--Marquardt \citep{press86}, a
$\chi^{2}$ minimization technique that employs a steepest descent
method. For the exponential and logarithmic laws, we omit the first
point of the intensity profiles from the fit as the function, or its
first derivative, becomes discontinuous at that point and makes our
fitting routine fail.

\begin{figure}[p]
\includegraphics[scale=0.80,bb=20 117 492 779]{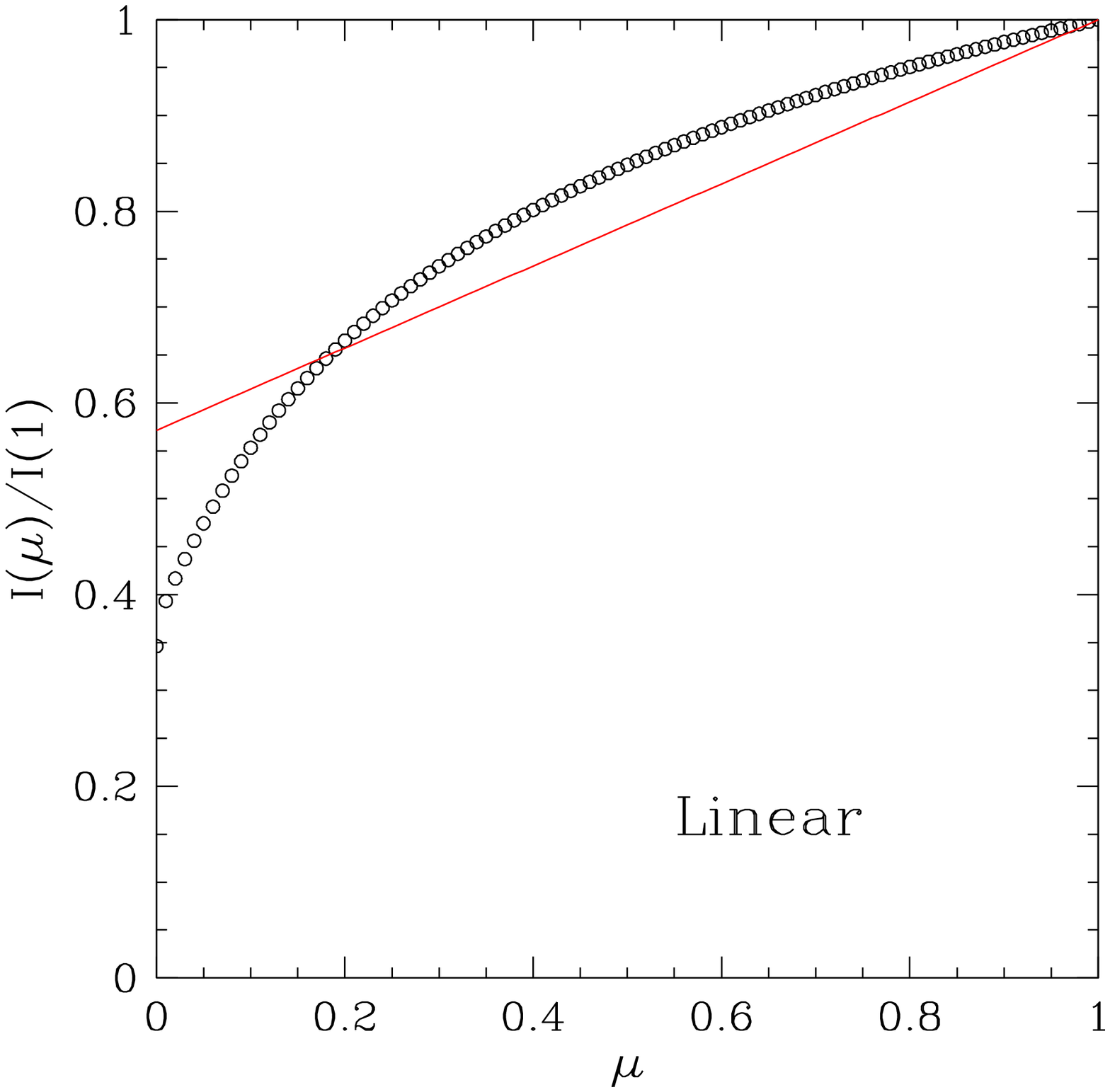}
\figcaption[f01.eps]{Fit to the intensity profile $I(\mu)/I(1)$
  computed for the LSST $g$-band filter and corresponding to a white
  dwarf model with \Te\ = 10,000 K and \logg\ = 8.0 using the linear
  limb-darkening law. The open circles correspond to the values of
  $I(\mu)/I(1)$ for 101 evenly spaced values of $\mu$ from 0 to 1 and
  the solid red line represents our best fit. \label{fg:lin}}
\end{figure}

\begin{figure}[p]
\includegraphics[scale=0.80,bb=20 117 492 779]{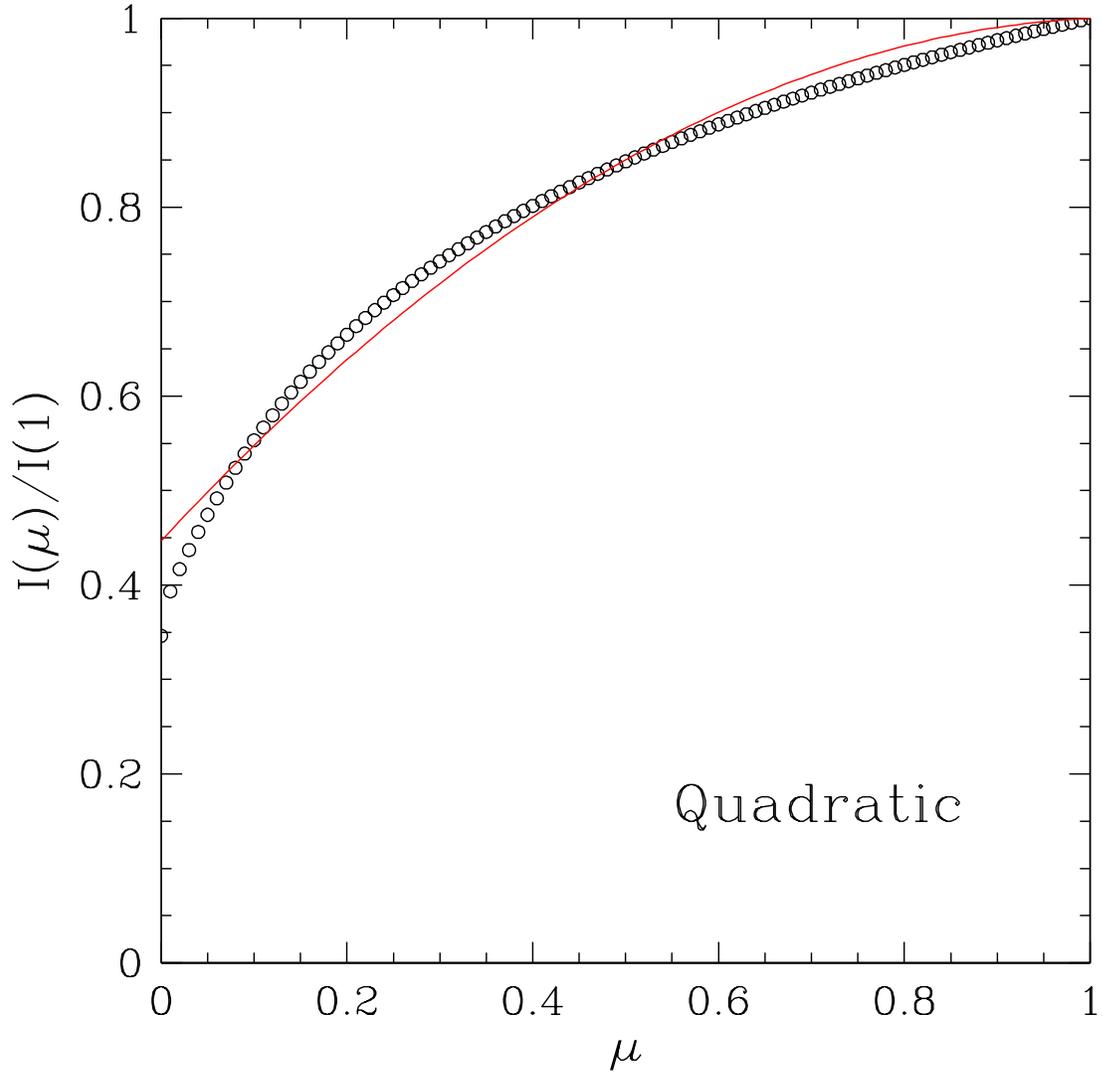}
\figcaption[f02.eps]{Same as Figure \ref{fg:lin} but using the
  quadratic limb-darkening law. \label{fg:quad}}
\end{figure}

\begin{figure}[p]
\includegraphics[scale=0.80,bb=20 117 492 779]{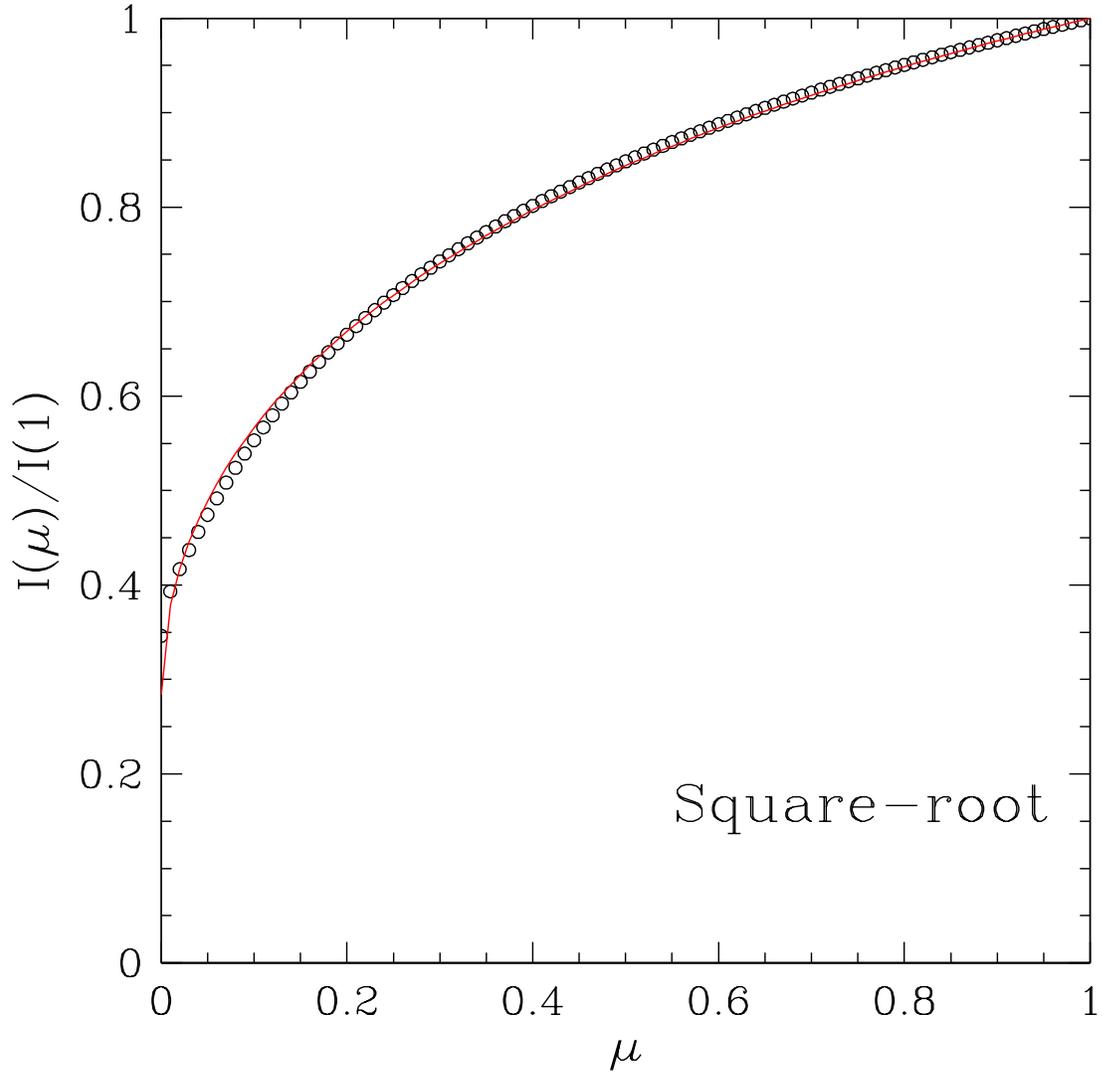}
\figcaption[f03.eps]{Same as Figure \ref{fg:lin} but using the
  square-root limb-darkening law. \label{fg:sqrt}}
\end{figure}

\begin{figure}[p]
\includegraphics[scale=0.80,bb=20 117 492 779]{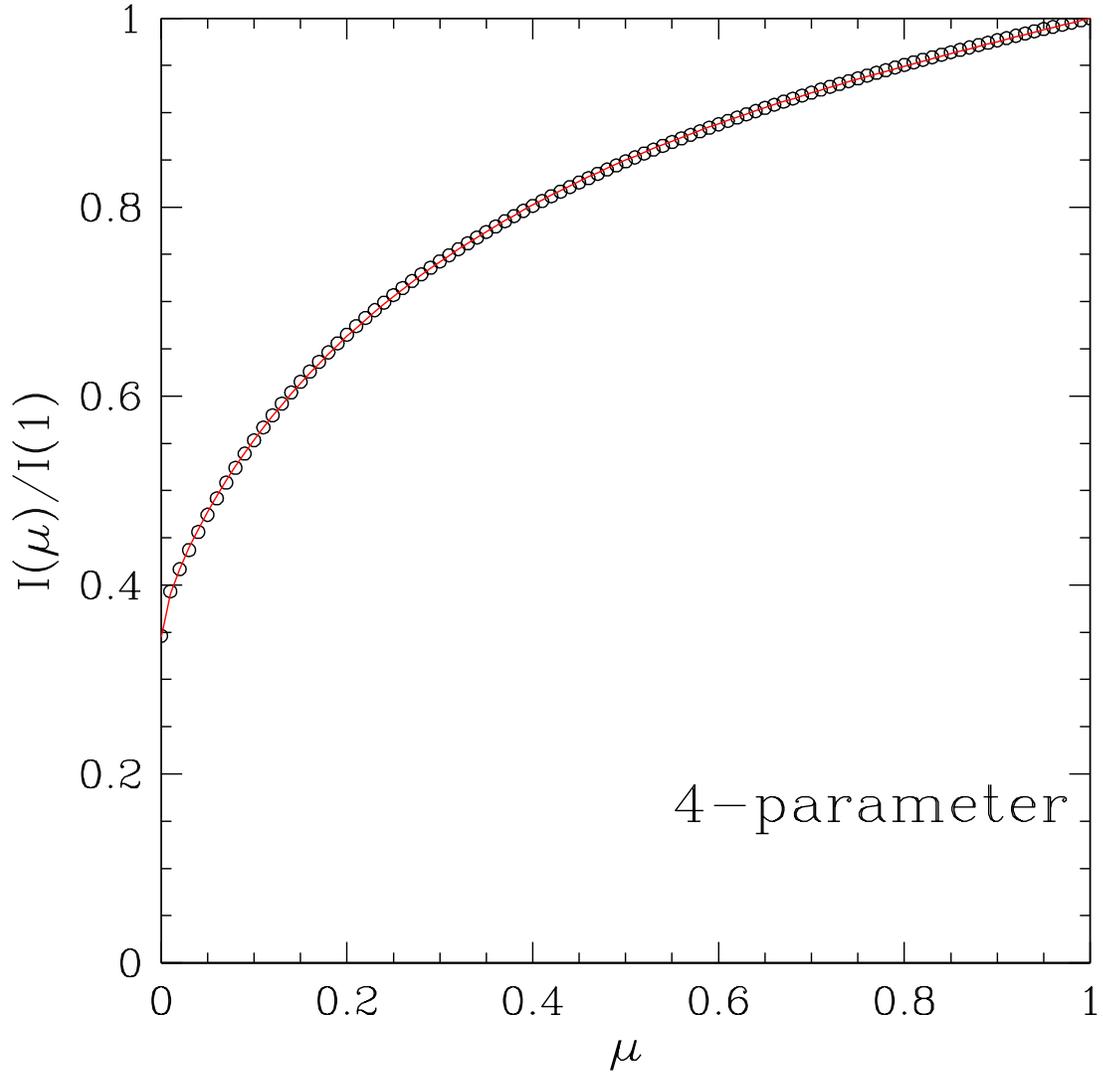}
\figcaption[f04.eps]{Same as Figure \ref{fg:lin} but using the Claret
  4-parameter limb-darkening law. \label{fg:4par}}
\end{figure}

We show in Figures \ref{fg:lin} through \ref{fg:4par} the result of
our fits using different limb-darkening laws for the intensity profile
of a typical white dwarf with \Te\ = 10,000~K, and \logg\ = 8.0 in the
LSST $g$-band filter. Figure \ref{fg:lin} demonstrates that
limb-darkening is not linear at all, as evidenced by how poorly the
linear law matches the intensity profile. Figure \ref{fg:quad} shows
that the quadratic form reproduces the intensity profile better,
though it overestimates the intensity close to the center of the disk
while underestimating it near the edge of the disk. Meanwhile, Figure
\ref{fg:sqrt} displays the square-root law and we note an even better
agreement than in the previous case though the fit does dip just below
the last point at $\mu$ = 0. Finally, Figure \ref{fg:4par} shows the
fit using the Claret 4-parameter limb-darkening law, which fits the
intensity profile extremely well.

Not surprisingly, the limb-darkening law with the most parameters, the
Claret 4-parameter limb-darkening law, produces by far the best fits
to all of the intensity profiles for our model grid. This is borne out
not only by visual inspection of the individual fits, but also by the
$\chi^{2}$ values which our fitting routine computes through the
covariance matrix. In all cases, the Claret 4-parameter law yields the
lowest $\chi^{2}$ values. However, many of the currently used
light-curve fitting codes do not include the Claret 4-parameter law,
which is why we present all seven enumerated limb-darkening laws in
this paper.

Table 1 presents the limb-darkening coefficients for the LSST $g$-band
filter for all seven limb-darkening laws and for our entire grid of
models with 627 different \Te\ and \logg\ values. The first two
columns provide \logg\ and \Te\ and the remaining 16 columns provide
the limb-darkening coefficients as defined in Section 3 and Equations
(1) through (7). This table is meant to serve as a guide for the form
of the Tables 2 through 11, published only in electronic form, listing
the limb-darkening coefficients for the 10 other filters considered in
this work. Finally, we also provide in Table 12 (in electronic form
only) exact intensity profiles computed for 16 evenly spaced values of
$\mu$ from 0 to 1. This will permit users the freedom to include
limb-darkening either in an exact fashion or by fitting the intensity
profiles with analytic expressions of their own choosing.

\section{CONVECTION}

\begin{figure}[p]
\includegraphics[scale=0.60,angle=-90]{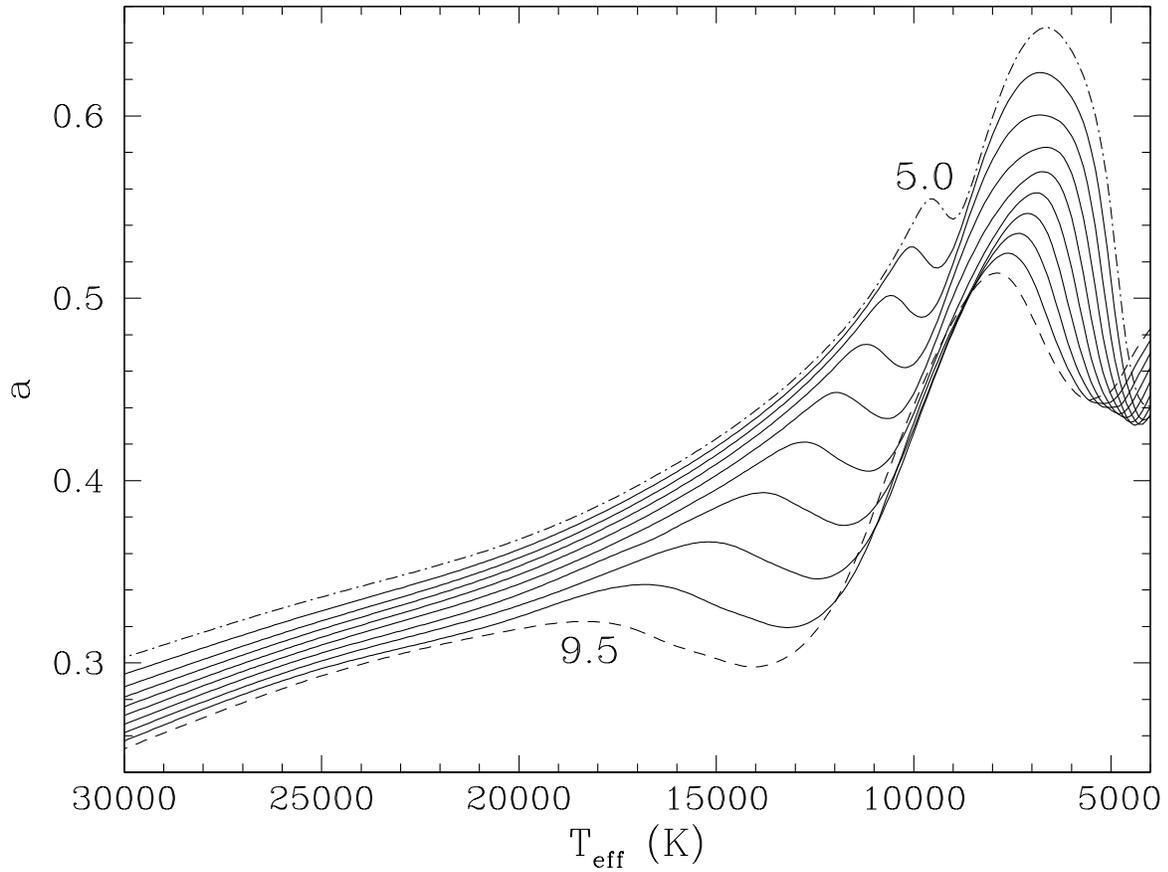}
\figcaption[f05.eps]{Plot of the coefficient $a$ from the linear
  limb-darkening law as a function of \Te, computed for the LSST
  $g$-band filter. The solid line curves correspond to models with
  \logg\ ranging from 5.0 (top, dash-dotted line) to 9.5 (bottom,
  dashed line) in steps of 0.5 dex, as indicated in the
  figure. \label{fg:coeffa}}
\end{figure}

\begin{figure}[p]
\includegraphics[scale=0.80,bb= 20 142 592 704]{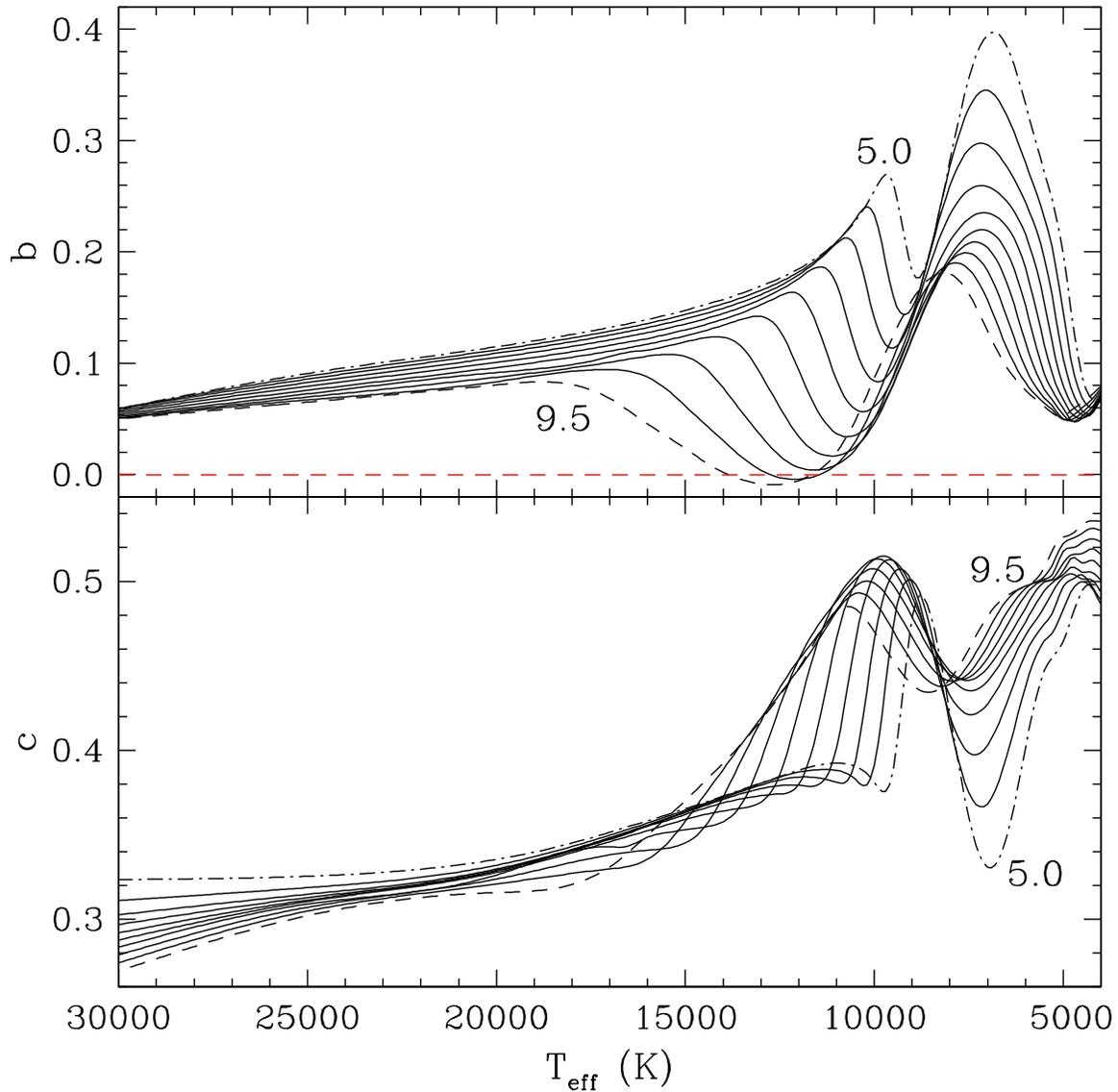}
\figcaption[f06.eps]{Plot of the coefficients $b$ (top panel) and $c$
  (bottom panel) from the quadratic limb-darkening law as a function
  of \Te, computed for the LSST $g$-band filter. The solid line curves
  correspond to models with \logg\ ranging from 5.0 (dash-dotted line)
  to 9.5 (dashed line) in steps of 0.5 dex, as indicated in the
  figure. The red dashed line in the top panel serves as a reference
  for where the value of $b$ becomes zero. We note that the value of
  $b$ becomes negative between $\sim$14,000~K and $\sim$11,000~K for
  the highest gravity models. \label{fg:coeffbc}}
\end{figure}

We explore in Figures \ref{fg:coeffa} and \ref{fg:coeffbc} how the
limb-darkening coefficients for the linear law ($a$) and those for the
quadratic law ($b,c$), computed for the LSST $g$-band filter, vary as
a function of \Te\ and \logg. The linear and quadratic laws are the
most widely used forms of limb-darkening.  Figure \ref{fg:coeffa}
demonstrates that the coefficient $a$ varies almost linearly as a
function of \Te\ until about $\sim$14,000~K. At cooler temperatures
the behavior becomes decidedly non-linear and the precise value of
\Te\ at which the departure from this quasi-linearity occurs decreases
as \logg\ decreases. In the top panel of Figure \ref{fg:coeffbc} the
variation of $b$ qualitatively resembles that of $a$. This is not
surprising as $b$ corresponds to the linear term in the quadratic
law. The main difference is that $b$ actually becomes negative between
$\sim$14,000~K and $\sim$11,000~K for the models with the highest
surface gravity in our grid. Similarly, the lower panel of Figure
\ref{fg:coeffbc} shows that $c$ also varies more or less linearly
until $\sim$15,000~K. The reason behind the sudden change at 15,000 K
becomes evident when we consider the effects of convection on
limb-darkening.

\begin{figure}[p]
\includegraphics[scale=0.70,angle=-90,bb=22 46 546 784]{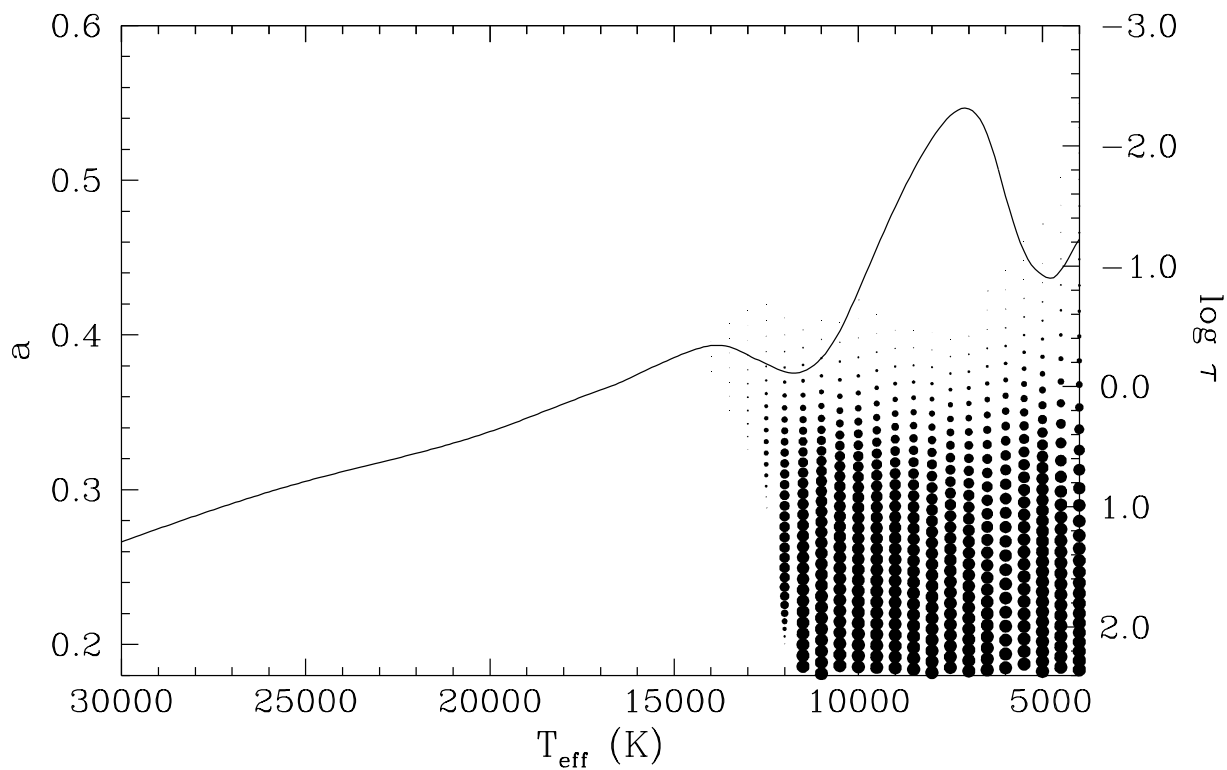}
\figcaption[f07.eps]{Plot of the coefficient $a$ of the linear
  limb-darkening law as a function of \Te\ (solid line) for \logg\ =
  8.0, computed for the LSST $g$-band filter. The scale on the
  right-hand side of the figure corresponds to the Rosseland optical
  depth, and the black dots represent the extent of the convection
  zone for each value of \Te\ in our model grid. The larger the dot,
  the higher the fraction of the total emergent flux that is
  transported via convection.\label{fg:conv}}
\end{figure}

As hydrogen-rich white dwarfs cool below $\sim$15,000~K, they develop
a convection zone which transports an increasingly large fraction of
the emergent flux. Figure \ref{fg:conv} shows the coefficient $a$ as a
function of \Te\ but only for \logg\ = 8.0. In addition, the scale on
the right-hand side of the figure gives the Rosseland optical depth.
The black dots correspond to the extent of the convection zone at
every \Te\ in our model grid. The size of the dots is proportional to
the fraction of the emergent flux that is being transported via
convection at that particular optical depth.  Figure \ref{fg:conv}
clearly demonstrates that the dip in the value of $a$ begins when
convective energy transport starts to become significant. The local
minimum that follows coincides with the atmosphere becoming completely
convective. Convection tends to flatten out the temperature gradient
in the atmosphere, which leads to less intense limb-darkening. This
translates to a shallower intensity profile and in the case of a
linear limb-darkening law, means that the slope, $a$, decreases.

\section{ECLIPSING WHITE DWARFS}

There are now 53 eclipsing white dwarfs known, 49 in white dwarf + M
dwarf binaries and four in double white dwarf systems. All four
eclipsing double white dwarf systems were discovered in the past two
years. With the ongoing and future transient surveys like the PTF,
Pan-STARRS, and the LSST, we expect that the number of eclipsing white
dwarf + M dwarf and double white dwarf systems will increase
significantly.

Due to its sensitivity and wide sky coverage, the LSST will find over
13 million white dwarfs with $r < 24.5$ mag, and over 50 million of
them in the final co-added catalog.  The LSST sample of short period
($P\leq0.3$ days) eclipsing binaries will be essentially complete to
$r \sim$ 24.5 mag \citep[see Figure 6.13 in][]{lsst09}. The
completeness drops to about 80\% for $P\approx1$~day binaries. About
22\% of field white dwarfs have stellar companions \citep[mostly
  low-mass main-sequence stars;][]{farihi05}, while 3.2\% of white
dwarf + main-sequence binaries are eclipsing systems with $P<1$~day
\citep{parsons13}. Hence, the frequency of eclipsing white dwarfs is
likely $\sim$0.7\%. Given the sheer number of white dwarfs imaged by
the LSST, the LSST may find $\sim$$10^5$ eclipsing white dwarfs with
stellar mass companions.  Our limb-darkening calculations will be an
essential part of the detailed analysis of all of these systems.

\section{TRANSITING EXOPLANETS}

\citet{agol11} has explored the possibility of finding Earth-sized
planets in the habitable zones around white dwarfs. If such systems
were to be viewed edge-on, even an Earth-sized planet would produce a
significant dip in the light-curve due to the comparable sizes of the
two objects. If every white dwarf has an Earth-like planet in its
habitable zone ($\sim$0.02~AU), we would expect to see 1\% of them
eclipse their stars \citep{agol11}. With 13 million white dwarfs in
its photometric catalog, the LSST will constrain the frequency of
planets in the habitable zone around white dwarfs to unprecedented
accuracy.

The detection rate of eclipsing systems falls off with increasing
orbital period. For example, the detection rate of a planet orbiting
at 1~AU from an average size white dwarf will be $\sim$10\%.
\citet{dstef10} estimate a transit probability of 0.02\% for a planet
at 1~AU from an average size white dwarf. Given the $\sim$10\%
detection rate, the LSST may find at least a few hundred planets, if
they exist within 1~AU.

\begin{figure}[p]
\includegraphics[scale=0.60,angle=-90]{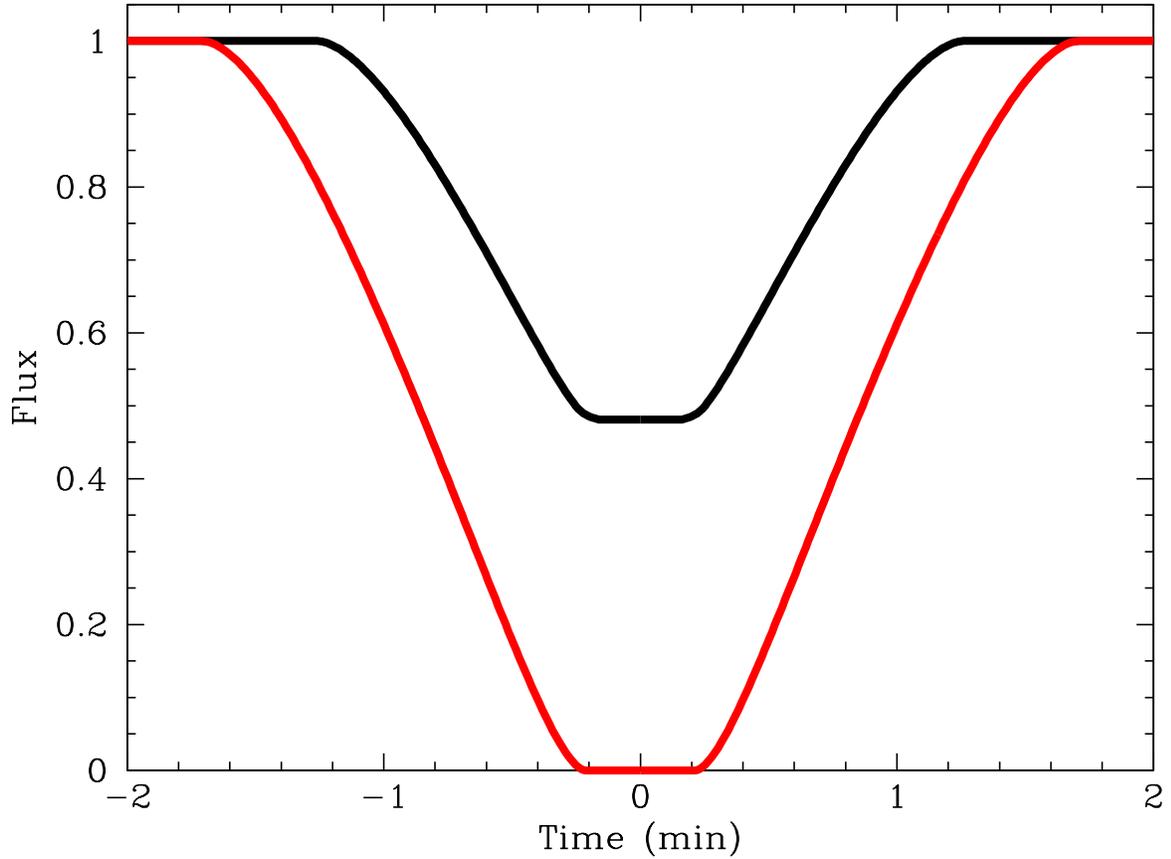}
\figcaption[f08.eps]{Simulation of eclipses in the $g$-band caused by
  a 1~\me\ planet (black) and a 10~\me\ (red) planet transiting in the
  habitable zone (0.01~AU) of a white dwarf with \Te\ = 5000~K and
  \logg\ = 8.0 using the JKTEBOP code. The limb-darkening coefficients
  from the quadratic law, appropriate for the chosen atmospheric
  parameters, were used (i.e., $b=0.0536$,
  $c=0.5110$).\label{fg:eclipse}}
\end{figure}

Here we model transiting Earths around white dwarfs using the {\sc
  jktebop} code\footnote{{\sc jktebop} is written in {\sc fortran77}
  and the source code is available at
  http://www.astro.keele.ac.uk/~jkt/} \citep{south04} which is a
modified version of the {\sc ebop} program
\citep{popper81,etzel81}. We consider a relatively cool white dwarf
with \Te\ =~5000~K and \logg\ =~8.0 as the host star and model the
effects of limb-darkening using the quadratic law and the appropriate
coefficients from our calculations.  Figure \ref{fg:eclipse} shows the
hypothetical light-curves of the aforementioned white dwarf as it is
eclipsed by a planet with $M=1$~\me\ and another with
$M=10$~\me\ orbiting in the habitable zone of the white dwarf that
corresponds to a distance of $\sim$0.01~AU. In the case of the
1~\me\ planet, the eclipse produces a reduction in flux of $\sim$50\%
whereas the potential 10~\me\ super-Earth could effectively block out
the white dwarf completely. Such systems would be relatively easy to
detect and a proper modeling of limb-darkening will allow for
accurate, and model-independent, measurements of the masses and sizes
of both the star and the planet.

\section{CONCLUSIONS}

The number of eclipsing white dwarf binaries known has increased
dramatically in the last two years. This emphasizes the need for
accurate limb-darkening coefficients for white dwarfs in order to
analyze these systems as accurately as possible. This need will only
increase as we anticipate that endeavors such as the LSST will uncover
$\sim10^5$ eclipsing white dwarf binaries, and possibly a few hundred
transiting planetary systems.

We use our up-to-date grid of atmosphere models of hydrogen-rich white
dwarfs in order to compute limb-darkening coefficients for seven
different limb-darkening laws. We chose to cover two separate
photometric systems; the widely used Johnson-Kron-Cousins $UBVRI$
photometric system as well as the LSST $ugrizy$ system.  We present
all of our calculations in online tables, which can be easily adopted
in light curve fitting codes like JKTEBOP and PHOEBE \citep{prsa05}.
The studies of large numbers of eclipsing white dwarfs found in the
ongoing and future transient surveys will greatly benefit from our
calculations of limb-darkening coefficients. Finally, we note once
again that these calculations can be easily extended to any
atmospheric composition and any photometric system, or even to higher
temperatures, upon request.

\noindent We thank J. Southworth for maintaining a Web site which
served as an excellent resource about limb-darkening in general as
well as listing references for all of the individual limb-darkening
laws we consider. This work is funded in part by the NSERC Canada and
by the Fund FRQ-NT (Qu\'ebec).

\clearpage

\bibliographystyle{apj}
\bibliography{biblio}

\clearpage

\clearpage

\clearpage

\clearpage

\clearpage

\clearpage

\clearpage

\clearpage

\clearpage
\begin{deluxetable}{cc|r|rr|rr|rr|rr|rrr|rrrr}
\rotate
\tabletypesize{\tiny}
\tablecolumns{18}
\tablewidth{0pt}
\tablecaption{Limb-darkening Coefficients for the LSST $g$-band}
\tablehead{
\colhead{\logg} &
\colhead{\Te} &
\colhead{Linear} &
\multicolumn{2}{c}{Quadratic} &
\multicolumn{2}{c}{Square-root} &
\multicolumn{2}{c}{Logarithmic} &
\multicolumn{2}{c}{Exponential} &
\multicolumn{3}{c}{Sing 3-parameter} &
\multicolumn{4}{c}{Claret 4-parameter} \\
\colhead{} &
\colhead{(K)} &
\colhead{$a$} &
\colhead{$b$} &
\colhead{$c$} &
\colhead{$d$} &
\colhead{$f$} &
\colhead{$g$} &
\colhead{$h$} &
\colhead{$j$} &
\colhead{$k$} &
\colhead{$l$} &
\colhead{$m$} &
\colhead{$n$} &
\colhead{$p$} &
\colhead{$q$} &
\colhead{$r$} &
\colhead{$s$}}
\startdata
5.00 &  4000 &  0.4407 &  0.0686 &  0.4937 & $-$0.2647 &  1.0005 &  0.6491 &  0.4760 &  0.3924 & $-$0.0046 &  3.1766 & $-$4.4234 &  1.9034 &  0.8417 &  0.4031 & $-$0.9647 &  0.4479 \\
5.00 &  4250 &  0.4437 &  0.0679 &  0.4985 & $-$0.2768 &  1.0219 &  0.6551 &  0.4830 &  0.3929 & $-$0.0048 &  3.2958 & $-$4.6532 &  2.0231 &  0.9980 &  0.0075 & $-$0.5525 &  0.2973 \\
5.00 &  4500 &  0.4612 &  0.0887 &  0.4943 & $-$0.2430 &  0.9988 &  0.6699 &  0.4760 &  0.4139 & $-$0.0046 &  3.1966 & $-$4.4206 &  1.9013 &  0.7583 &  0.6979 & $-$1.3047 &  0.5900 \\
5.00 &  4750 &  0.4990 &  0.1369 &  0.4804 & $-$0.1712 &  0.9506 &  0.7005 &  0.4575 &  0.4573 & $-$0.0041 &  3.0036 & $-$3.9755 &  1.6740 &  0.4742 &  1.4408 & $-$2.0264 &  0.8537 \\
5.00 &  5000 &  0.5481 &  0.1980 &  0.4645 & $-$0.0802 &  0.8911 &  0.7407 &  0.4347 &  0.5132 & $-$0.0036 &  2.7234 & $-$3.3297 &  1.3399 &  0.1735 &  2.1518 & $-$2.6169 &  1.0399 \\
5.00 &  5250 &  0.5848 &  0.2412 &  0.4559 & $-$0.0089 &  0.8421 &  0.7709 &  0.4182 &  0.5560 & $-$0.0032 &  2.4137 & $-$2.6298 &  0.9692 & $-$0.1108 &  2.7789 & $-$3.0852 &  1.1608 \\
5.00 &  5500 &  0.6086 &  0.2716 &  0.4471 &  0.0416 &  0.8043 &  0.7890 &  0.4036 &  0.5836 & $-$0.0029 &  2.1625 & $-$2.0790 &  0.6795 & $-$0.2026 &  2.8300 & $-$2.9113 &  1.0298 \\
5.00 &  6000 &  0.6356 &  0.3332 &  0.4013 &  0.1447 &  0.6963 &  0.7947 &  0.3524 &  0.6172 & $-$0.0024 &  1.6952 & $-$1.1668 &  0.2310 & $-$0.1656 &  2.2409 & $-$1.8474 &  0.5174 \\
\enddata
\label{tab:g}
\end{deluxetable}

\clearpage
\begin{deluxetable}{cc|r|rr|rr|rr|rr|rrr|rrrr}
\rotate
\tabletypesize{\tiny}
\tablecolumns{18}
\tablewidth{0pt}
\tablecaption{Limb-darkening Coefficients for the LSST $u$-band}
\tablehead{
\colhead{\logg} &
\colhead{\Te} &
\colhead{Linear} &
\multicolumn{2}{c}{Quadratic} &
\multicolumn{2}{c}{Square-root} &
\multicolumn{2}{c}{Logarithmic} &
\multicolumn{2}{c}{Exponential} &
\multicolumn{3}{c}{Sing 3-parameter} &
\multicolumn{4}{c}{Claret 4-parameter} \\
\colhead{} &
\colhead{(K)} &
\colhead{$a$} &
\colhead{$b$} &
\colhead{$c$} &
\colhead{$d$} &
\colhead{$f$} &
\colhead{$g$} &
\colhead{$h$} &
\colhead{$j$} &
\colhead{$k$} &
\colhead{$l$} &
\colhead{$m$} &
\colhead{$n$} &
\colhead{$p$} &
\colhead{$q$} &
\colhead{$r$} &
\colhead{$s$}}
\startdata
5.00 &  4000 &  0.5731 &  0.2169 &  0.4726 & $-$0.0863 &  0.9352 &  0.7558 &  0.4420 &  0.5163 & $-$0.0044 &  2.8267 & $-$3.4723 &  1.4091 &  0.9965 & $-$0.4566 &  0.6222 & $-$0.3140 \\
5.00 &  4250 &  0.5763 &  0.2207 &  0.4719 & $-$0.0817 &  0.9334 &  0.7580 &  0.4403 &  0.5201 & $-$0.0043 &  2.8600 & $-$3.5365 &  1.4446 &  0.8932 & $-$0.0833 &  0.1338 & $-$0.1001 \\
5.00 &  4500 &  0.5992 &  0.2740 &  0.4313 &  0.0207 &  0.8205 &  0.7643 &  0.3995 &  0.5532 & $-$0.0035 &  2.4410 & $-$2.7204 &  1.0429 &  0.3803 &  1.1877 & $-$1.1575 &  0.3852 \\
5.00 &  4750 &  0.6382 &  0.3419 &  0.3932 &  0.1335 &  0.7159 &  0.7876 &  0.3608 &  0.6017 & $-$0.0028 &  2.0777 & $-$1.9803 &  0.6799 & $-$0.0853 &  2.3587 & $-$2.3306 &  0.8274 \\
5.00 &  5000 &  0.6788 &  0.3980 &  0.3726 &  0.2180 &  0.6536 &  0.8194 &  0.3392 &  0.6481 & $-$0.0023 &  1.8361 & $-$1.4452 &  0.4106 & $-$0.3579 &  3.0153 & $-$2.9158 &  1.0295 \\
5.00 &  5250 &  0.7036 &  0.4170 &  0.3802 &  0.2515 &  0.6413 &  0.8459 &  0.3430 &  0.6763 & $-$0.0020 &  1.6735 & $-$1.0349 &  0.1806 & $-$0.6576 &  3.8405 & $-$3.7372 &  1.3179 \\
5.00 &  5500 &  0.7190 &  0.4166 &  0.4012 &  0.2449 &  0.6725 &  0.8690 &  0.3612 &  0.6906 & $-$0.0021 &  1.6744 & $-$0.9497 &  0.1134 & $-$0.6592 &  3.8465 & $-$3.6584 &  1.2534 \\
5.00 &  6000 &  0.7350 &  0.4249 &  0.4114 &  0.2539 &  0.6823 &  0.8878 &  0.3678 &  0.7060 & $-$0.0022 &  1.5589 & $-$0.6493 & $-$0.0595 & $-$0.5172 &  3.2633 & $-$2.7747 &  0.8350 \\
\enddata
\label{tab:u}
\end{deluxetable}

\clearpage
\begin{deluxetable}{cc|r|rr|rr|rr|rr|rrr|rrrr}
\rotate
\tabletypesize{\tiny}
\tablecolumns{18}
\tablewidth{0pt}
\tablecaption{Limb-darkening Coefficients for the LSST $r$-band}
\tablehead{
\colhead{\logg} &
\colhead{\Te} &
\colhead{Linear} &
\multicolumn{2}{c}{Quadratic} &
\multicolumn{2}{c}{Square-root} &
\multicolumn{2}{c}{Logarithmic} &
\multicolumn{2}{c}{Exponential} &
\multicolumn{3}{c}{Sing 3-parameter} &
\multicolumn{4}{c}{Claret 4-parameter} \\
\colhead{} &
\colhead{(K)} &
\colhead{$a$} &
\colhead{$b$} &
\colhead{$c$} &
\colhead{$d$} &
\colhead{$f$} &
\colhead{$g$} &
\colhead{$h$} &
\colhead{$j$} &
\colhead{$k$} &
\colhead{$l$} &
\colhead{$m$} &
\colhead{$n$} &
\colhead{$p$} &
\colhead{$q$} &
\colhead{$r$} &
\colhead{$s$}}
\startdata
5.00 & 4000 & 0.3569 & 0.0055 & 0.4663 & $-$0.3365 & 0.9834 & 0.5569 &
0.4592 & 0.3043 & $-$0.0049 & 3.2720 & $-$4.8684 & 2.1720 & 1.2019 &
$-$0.6883 & 0.0703 & 0.0936 \\ 5.00 & 4250 & 0.3608 & 0.0061 & 0.4705
& $-$0.3464 & 1.0030 & 0.5633 & 0.4649 & 0.3056 & $-$0.0051 & 3.3663 &
$-$5.0460 & 2.2639 & 1.3969 & $-$1.2366 & 0.6941 & $-$0.1517 \\ 5.00 &
4500 & 0.3765 & 0.0227 & 0.4694 & $-$0.3162 & 0.9825 & 0.5774 & 0.4605
& 0.3252 & $-$0.0048 & 3.2575 & $-$4.7890 & 2.1258 & 1.1006 &
$-$0.3690 & $-$0.2666 & 0.2226 \\ 5.00 & 4750 & 0.4097 & 0.0607 &
0.4630 & $-$0.2564 & 0.9448 & 0.6061 & 0.4487 & 0.3639 & $-$0.0044 &
3.0684 & $-$4.3421 & 1.8900 & 0.7648 & 0.5480 & $-$1.1989 & 0.5672 \\
5.00 & 5000 & 0.4531 & 0.1128 & 0.4516 & $-$0.1730 & 0.8881 & 0.6420 &
0.4293 & 0.4146 & $-$0.0038 & 2.7613 & $-$3.6406 & 1.5213 & 0.3858 &
1.4902 & $-$2.0555 & 0.8542 \\ 5.00 & 5250 & 0.4870 & 0.1570 & 0.4378
& $-$0.0951 & 0.8257 & 0.6669 & 0.4071 & 0.4555 & $-$0.0033 & 2.3906 &
$-$2.8325 & 1.0971 & 0.0717 & 2.1544 & $-$2.5380 & 0.9732 \\ 5.00 &
5500 & 0.5089 & 0.1932 & 0.4189 & $-$0.0309 & 0.7657 & 0.6787 & 0.3821
& 0.4825 & $-$0.0029 & 2.0637 & $-$2.1542 & 0.7485 & $-$0.0434 &
2.2066 & $-$2.3324 & 0.8235 \\ 5.00 & 6000 & 0.5303 & 0.2599 & 0.3588
& 0.0841 & 0.6329 & 0.6732 & 0.3180 & 0.5114 & $-$0.0023 & 1.5556 &
$-$1.2059 & 0.2947 & $-$0.0116 & 1.5939 & $-$1.2536 & 0.3148 \\
\enddata
\label{tab:r}
\end{deluxetable}

\clearpage
\begin{deluxetable}{cc|r|rr|rr|rr|rr|rrr|rrrr}
\rotate
\tabletypesize{\tiny}
\tablecolumns{18}
\tablewidth{0pt}
\tablecaption{Limb-darkening Coefficients for the LSST $i$-band}
\tablehead{
\colhead{\logg} &
\colhead{\Te} &
\colhead{Linear} &
\multicolumn{2}{c}{Quadratic} &
\multicolumn{2}{c}{Square-root} &
\multicolumn{2}{c}{Logarithmic} &
\multicolumn{2}{c}{Exponential} &
\multicolumn{3}{c}{Sing 3-parameter} &
\multicolumn{4}{c}{Claret 4-parameter} \\
\colhead{} &
\colhead{(K)} &
\colhead{$a$} &
\colhead{$b$} &
\colhead{$c$} &
\colhead{$d$} &
\colhead{$f$} &
\colhead{$g$} &
\colhead{$h$} &
\colhead{$j$} &
\colhead{$k$} &
\colhead{$l$} &
\colhead{$m$} &
\colhead{$n$} &
\colhead{$p$} &
\colhead{$q$} &
\colhead{$r$} &
\colhead{$s$}}
\startdata
5.00 &  4000 &  0.3045 & $-$0.0201 &  0.4306 & $-$0.3564 &  0.9374 &  0.4912 &  0.4297 &  0.2503 & $-$0.0049 &  3.1973 & $-$4.9144 &  2.2325 &  1.5460 & $-$1.8969 &  1.4383 & $-$0.4409 \\
5.00 &  4250 &  0.3086 & $-$0.0173 &  0.4324 & $-$0.3566 &  0.9436 &  0.4963 &  0.4317 &  0.2537 & $-$0.0050 &  3.2251 & $-$4.9591 &  2.2550 &  1.5968 & $-$2.0365 &  1.6024 & $-$0.5064 \\
5.00 &  4500 &  0.3227 & $-$0.0013 &  0.4299 & $-$0.3208 &  0.9128 &  0.5076 &  0.4248 &  0.2732 & $-$0.0046 &  3.0644 & $-$4.6014 &  2.0637 &  1.1696 & $-$0.7893 &  0.2045 &  0.0412 \\
5.00 &  4750 &  0.3519 &  0.0304 &  0.4265 & $-$0.2683 &  0.8797 &  0.5335 &  0.4159 &  0.3077 & $-$0.0041 &  2.8804 & $-$4.1655 &  1.8308 &  0.8081 &  0.2177 & $-$0.8448 &  0.4333 \\
5.00 &  5000 &  0.3902 &  0.0754 &  0.4177 & $-$0.1935 &  0.8278 &  0.5653 &  0.3992 &  0.3531 & $-$0.0036 &  2.5823 & $-$3.4867 &  1.4718 &  0.4193 &  1.2007 & $-$1.7637 &  0.7467 \\
5.00 &  5250 &  0.4212 &  0.1178 &  0.4025 & $-$0.1167 &  0.7630 &  0.5869 &  0.3758 &  0.3912 & $-$0.0031 &  2.2065 & $-$2.6774 &  1.0484 &  0.1034 &  1.8659 & $-$2.2526 &  0.8696 \\
5.00 &  5500 &  0.4410 &  0.1534 &  0.3815 & $-$0.0529 &  0.7006 &  0.5958 &  0.3493 &  0.4160 & $-$0.0027 &  1.8813 & $-$2.0124 &  0.7090 & $-$0.0006 &  1.8833 & $-$2.0148 &  0.7100 \\
5.00 &  6000 &  0.4585 &  0.2159 &  0.3219 &  0.0552 &  0.5720 &  0.5870 &  0.2866 &  0.4406 & $-$0.0021 &  1.4118 & $-$1.1515 &  0.3021 &  0.0414 &  1.2753 & $-$0.9812 &  0.2304 \\
\enddata
\label{tab:i}
\end{deluxetable}

\include{tab05}
\clearpage
\begin{deluxetable}{cc|r|rr|rr|rr|rr|rrr|rrrr}
\rotate
\tabletypesize{\tiny}
\tablecolumns{18}
\tablewidth{0pt}
\tablecaption{Limb-darkening Coefficients for the LSST $y$-band}
\tablehead{
\colhead{\logg} &
\colhead{\Te} &
\colhead{Linear} &
\multicolumn{2}{c}{Quadratic} &
\multicolumn{2}{c}{Square-root} &
\multicolumn{2}{c}{Logarithmic} &
\multicolumn{2}{c}{Exponential} &
\multicolumn{3}{c}{Sing 3-parameter} &
\multicolumn{4}{c}{Claret 4-parameter} \\
\colhead{} &
\colhead{(K)} &
\colhead{$a$} &
\colhead{$b$} &
\colhead{$c$} &
\colhead{$d$} &
\colhead{$f$} &
\colhead{$g$} &
\colhead{$h$} &
\colhead{$j$} &
\colhead{$k$} &
\colhead{$l$} &
\colhead{$m$} &
\colhead{$n$} &
\colhead{$p$} &
\colhead{$q$} &
\colhead{$r$} &
\colhead{$s$}}
\startdata
5.00 &  4000 &  0.2494 & $-$0.0352 &  0.3776 & $-$0.3457 &  0.8441 &  0.4146 &  0.3809 &  0.1975 & $-$0.0047 &  2.9305 & $-$4.6105 &  2.1208 &  1.7036 & $-$2.6830 &  2.3898 & $-$0.8252 \\
5.00 &  4250 &  0.2529 & $-$0.0310 &  0.3767 & $-$0.3374 &  0.8373 &  0.4174 &  0.3791 &  0.2021 & $-$0.0046 &  2.8994 & $-$4.5410 &  2.0842 &  1.6193 & $-$2.4362 &  2.1129 & $-$0.7161 \\
5.00 &  4500 &  0.2652 & $-$0.0169 &  0.3742 & $-$0.3021 &  0.8046 &  0.4269 &  0.3721 &  0.2201 & $-$0.0041 &  2.7294 & $-$4.1670 &  1.8839 &  1.1458 & $-$1.0461 &  0.5412 & $-$0.0975 \\
5.00 &  4750 &  0.2906 &  0.0079 &  0.3751 & $-$0.2608 &  0.7821 &  0.4509 &  0.3679 &  0.2501 & $-$0.0038 &  2.5840 & $-$3.8088 &  1.6890 &  0.8066 & $-$0.0738 & $-$0.4944 &  0.2941 \\
5.00 &  5000 &  0.3238 &  0.0434 &  0.3720 & $-$0.2012 &  0.7446 &  0.4803 &  0.3577 &  0.2892 & $-$0.0033 &  2.3422 & $-$3.2431 &  1.3855 &  0.4447 &  0.8770 & $-$1.4159 &  0.6165 \\
5.00 &  5250 &  0.3508 &  0.0783 &  0.3615 & $-$0.1366 &  0.6912 &  0.5000 &  0.3396 &  0.3225 & $-$0.0028 &  2.0157 & $-$2.5331 &  1.0112 &  0.1332 &  1.5767 & $-$1.9857 &  0.7809 \\
5.00 &  5500 &  0.3682 &  0.1081 &  0.3450 & $-$0.0827 &  0.6396 &  0.5087 &  0.3182 &  0.3443 & $-$0.0025 &  1.7356 & $-$1.9552 &  0.7145 &  0.0322 &  1.6294 & $-$1.8228 &  0.6588 \\
5.00 &  6000 &  0.3835 &  0.1608 &  0.2955 &  0.0091 &  0.5310 &  0.5019 &  0.2656 &  0.3656 & $-$0.0020 &  1.3272 & $-$1.2004 &  0.3551 &  0.0594 &  1.1313 & $-$0.9562 &  0.2523 \\
\enddata
\label{tab:y}
\end{deluxetable}

\clearpage
\begin{deluxetable}{cc|r|rr|rr|rr|rr|rrr|rrrr}
\rotate
\tabletypesize{\tiny}
\tablecolumns{18}
\tablewidth{0pt}
\tablecaption{Limb-darkening Coefficients for the Johnson--Kron--Cousins $U$-band}
\tablehead{
\colhead{\logg} &
\colhead{\Te} &
\colhead{Linear} &
\multicolumn{2}{c}{Quadratic} &
\multicolumn{2}{c}{Square-root} &
\multicolumn{2}{c}{Logarithmic} &
\multicolumn{2}{c}{Exponential} &
\multicolumn{3}{c}{Sing 3-parameter} &
\multicolumn{4}{c}{Claret 4-parameter} \\
\colhead{} &
\colhead{(K)} &
\colhead{$a$} &
\colhead{$b$} &
\colhead{$c$} &
\colhead{$d$} &
\colhead{$f$} &
\colhead{$g$} &
\colhead{$h$} &
\colhead{$j$} &
\colhead{$k$} &
\colhead{$l$} &
\colhead{$m$} &
\colhead{$n$} &
\colhead{$p$} &
\colhead{$q$} &
\colhead{$r$} &
\colhead{$s$}}
\startdata
5.00 & 4000 & 0.5736 & 0.2178 & 0.4720 & $-$0.0851 & 0.9344 & 0.7701 &
0.4430 & 0.5311 & $-$0.0043 & 2.8251 & $-$3.4690 & 1.4078 & 0.9977 &
$-$0.4622 & 0.6304 & $-$0.3173 \\ 5.00 & 4250 & 0.5772 & 0.2221 &
0.4710 & $-$0.0798 & 0.9318 & 0.7735 & 0.4426 & 0.5354 & $-$0.0042 &
2.8558 & $-$3.5283 & 1.4410 & 0.8971 & $-$0.1003 & 0.1580 &
$-$0.1104 \\ 5.00 & 4500 & 0.6006 & 0.2773 & 0.4288 & 0.0253 & 0.8159
& 0.7771 & 0.3949 & 0.5693 & $-$0.0034 & 2.4291 & $-$2.6993 & 1.0339 &
0.3895 & 1.1456 & $-$1.0987 & 0.3603 \\ 5.00 & 4750 & 0.6413 & 0.3490
& 0.3878 & 0.1441 & 0.7052 & 0.7988 & 0.3485 & 0.6202 & $-$0.0026 &
2.0487 & $-$1.9275 & 0.6567 & $-$0.0809 & 2.3152 & $-$2.2599 &
0.7966 \\ 5.00 & 5000 & 0.6827 & 0.4069 & 0.3659 & 0.2316 & 0.6397 &
0.8294 & 0.3216 & 0.6682 & $-$0.0021 & 1.7943 & $-$1.3676 & 0.3752 &
$-$0.3603 & 2.9817 & $-$2.8483 & 0.9983 \\ 5.00 & 5250 & 0.7072 &
0.4257 & 0.3735 & 0.2652 & 0.6269 & 0.8548 & 0.3229 & 0.6968 &
$-$0.0018 & 1.6274 & $-$0.9483 & 0.1404 & $-$0.6699 & 3.8346 &
$-$3.7009 & 1.2988 \\ 5.00 & 5500 & 0.7224 & 0.4243 & 0.3954 & 0.2573
& 0.6597 & 0.8780 & 0.3412 & 0.7111 & $-$0.0019 & 1.6332 & $-$0.8717 &
0.0769 & $-$0.6739 & 3.8538 & $-$3.6410 & 1.2424 \\ 5.00 & 6000 &
0.7384 & 0.4313 & 0.4074 & 0.2636 & 0.6734 & 0.8974 & 0.3485 & 0.7265
& $-$0.0020 & 1.5300 & $-$0.5921 & $-$0.0865 & $-$0.5338 & 3.2890 &
$-$2.7857 & 0.8366 \\
\enddata
\label{tab:U}
\end{deluxetable}

\clearpage
\begin{deluxetable}{cc|r|rr|rr|rr|rr|rrr|rrrr}
\rotate
\tabletypesize{\tiny}
\tablecolumns{18}
\tablewidth{0pt}
\tablecaption{Limb-darkening Coefficients for the Johnson--Kron--Cousins $B$-band}
\tablehead{
\colhead{\logg} &
\colhead{\Te} &
\colhead{Linear} &
\multicolumn{2}{c}{Quadratic} &
\multicolumn{2}{c}{Square-root} &
\multicolumn{2}{c}{Logarithmic} &
\multicolumn{2}{c}{Exponential} &
\multicolumn{3}{c}{Sing 3-parameter} &
\multicolumn{4}{c}{Claret 4-parameter} \\
\colhead{} &
\colhead{(K)} &
\colhead{$a$} &
\colhead{$b$} &
\colhead{$c$} &
\colhead{$d$} &
\colhead{$f$} &
\colhead{$g$} &
\colhead{$h$} &
\colhead{$j$} &
\colhead{$k$} &
\colhead{$l$} &
\colhead{$m$} &
\colhead{$n$} &
\colhead{$p$} &
\colhead{$q$} &
\colhead{$r$} &
\colhead{$s$}}
\startdata
5.00 &  4000 &  0.3979 &  0.0323 &  0.4850 & $-$0.3071 &  1.0000 &  0.6043 &  0.4726 &  0.3475 & $-$0.0047 &  3.2515 & $-$4.6911 &  2.0571 &  0.9580 &  0.0948 & $-$0.7545 &  0.4004 \\
5.00 &  4250 &  0.4008 &  0.0315 &  0.4899 & $-$0.3210 &  1.0237 &  0.6101 &  0.4796 &  0.3473 & $-$0.0050 &  3.3738 & $-$4.9277 &  2.1805 &  1.1805 & $-$0.5161 & $-$0.0768 &  0.1390 \\
5.00 &  4500 &  0.4170 &  0.0486 &  0.4889 & $-$0.2937 &  1.0081 &  0.6251 &  0.4760 &  0.3665 & $-$0.0048 &  3.2965 & $-$4.7352 &  2.0772 &  0.9646 &  0.1178 & $-$0.7712 &  0.4089 \\
5.00 &  4750 &  0.4527 &  0.0912 &  0.4796 & $-$0.2294 &  0.9674 &  0.6552 &  0.4617 &  0.4074 & $-$0.0044 &  3.1122 & $-$4.3006 &  1.8509 &  0.6667 &  0.9153 & $-$1.5609 &  0.6980 \\
5.00 &  5000 &  0.4992 &  0.1477 &  0.4664 & $-$0.1421 &  0.9095 &  0.6937 &  0.4410 &  0.4611 & $-$0.0038 &  2.8145 & $-$3.6169 &  1.4936 &  0.3207 &  1.7577 & $-$2.2990 &  0.9390 \\
5.00 &  5250 &  0.5354 &  0.1928 &  0.4545 & $-$0.0650 &  0.8516 &  0.7218 &  0.4206 &  0.5040 & $-$0.0033 &  2.4636 & $-$2.8404 &  1.0848 &  0.0164 &  2.4096 & $-$2.7731 &  1.0564 \\
5.00 &  5500 &  0.5585 &  0.2275 &  0.4392 & $-$0.0047 &  0.7989 &  0.7362 &  0.3993 &  0.5318 & $-$0.0030 &  2.1581 & $-$2.1949 &  0.7502 & $-$0.0970 &  2.4778 & $-$2.5936 &  0.9180 \\
5.00 &  6000 &  0.5829 &  0.2944 &  0.3828 &  0.1099 &  0.6709 &  0.7351 &  0.3381 &  0.5638 & $-$0.0024 &  1.6450 & $-$1.2203 &  0.2785 & $-$0.0702 &  1.8763 & $-$1.5085 &  0.3997 \\
\enddata
\label{tab:B}
\end{deluxetable}

\clearpage
\begin{deluxetable}{cc|r|rr|rr|rr|rr|rrr|rrrr}
\rotate
\tabletypesize{\tiny}
\tablecolumns{18}
\tablewidth{0pt}
\tablecaption{Limb-darkening Coefficients for the Johnson--Kron--Cousins $V$-band}
\tablehead{
\colhead{\logg} &
\colhead{\Te} &
\colhead{Linear} &
\multicolumn{2}{c}{Quadratic} &
\multicolumn{2}{c}{Square-root} &
\multicolumn{2}{c}{Logarithmic} &
\multicolumn{2}{c}{Exponential} &
\multicolumn{3}{c}{Sing 3-parameter} &
\multicolumn{4}{c}{Claret 4-parameter} \\
\colhead{} &
\colhead{(K)} &
\colhead{$a$} &
\colhead{$b$} &
\colhead{$c$} &
\colhead{$d$} &
\colhead{$f$} &
\colhead{$g$} &
\colhead{$h$} &
\colhead{$j$} &
\colhead{$k$} &
\colhead{$l$} &
\colhead{$m$} &
\colhead{$n$} &
\colhead{$p$} &
\colhead{$q$} &
\colhead{$r$} &
\colhead{$s$}}
\startdata
5.00 &  4000 &  0.4859 &  0.1123 &  0.4957 & $-$0.2128 &  0.9909 &  0.6937 &  0.4731 &  0.4396 & $-$0.0045 &  3.0729 & $-$4.1076 &  1.7303 &  0.8067 &  0.4146 & $-$0.7925 &  0.3352 \\
5.00 &  4250 &  0.4875 &  0.1105 &  0.5001 & $-$0.2232 &  1.0080 &  0.6980 &  0.4794 &  0.4396 & $-$0.0046 &  3.1800 & $-$4.3161 &  1.8388 &  0.8694 &  0.3153 & $-$0.7437 &  0.3354 \\
5.00 &  4500 &  0.5044 &  0.1348 &  0.4903 & $-$0.1796 &  0.9701 &  0.7097 &  0.4664 &  0.4613 & $-$0.0043 &  3.0296 & $-$3.9928 &  1.6735 &  0.5598 &  1.1851 & $-$1.6926 &  0.7055 \\
5.00 &  4750 &  0.5416 &  0.1865 &  0.4711 & $-$0.1017 &  0.9125 &  0.7376 &  0.4431 &  0.5046 & $-$0.0038 &  2.8214 & $-$3.5305 &  1.4420 &  0.2677 &  1.9394 & $-$2.4306 &  0.9791 \\
5.00 &  5000 &  0.5903 &  0.2490 &  0.4528 & $-$0.0105 &  0.8522 &  0.7768 &  0.4187 &  0.5597 & $-$0.0033 &  2.5620 & $-$2.9356 &  1.1380 &  0.0104 &  2.5279 & $-$2.8930 &  1.1201 \\
5.00 &  5250 &  0.6255 &  0.2880 &  0.4477 &  0.0525 &  0.8127 &  0.8072 &  0.4063 &  0.6003 & $-$0.0029 &  2.2971 & $-$2.3230 &  0.8112 & $-$0.2539 &  3.1338 & $-$3.3663 &  1.2502 \\
5.00 &  5500 &  0.6484 &  0.3125 &  0.4456 &  0.0906 &  0.7911 &  0.8274 &  0.3991 &  0.6258 & $-$0.0027 &  2.1075 & $-$1.8851 &  0.5759 & $-$0.3161 &  3.1491 & $-$3.1841 &  1.1226 \\
5.00 &  6000 &  0.6748 &  0.3647 &  0.4115 &  0.1764 &  0.7070 &  0.8375 &  0.3595 &  0.6574 & $-$0.0023 &  1.7054 & $-$1.0797 &  0.1736 & $-$0.2563 &  2.5500 & $-$2.1330 &  0.6169 \\
\enddata
\label{tab:V}
\end{deluxetable}

\clearpage
\begin{deluxetable}{cc|r|rr|rr|rr|rr|rrr|rrrr}
\rotate
\tabletypesize{\tiny}
\tablecolumns{18}
\tablewidth{0pt}
\tablecaption{Limb-darkening Coefficients for the Johnson--Kron--Cousins $R$-band}
\tablehead{
\colhead{\logg} &
\colhead{\Te} &
\colhead{Linear} &
\multicolumn{2}{c}{Quadratic} &
\multicolumn{2}{c}{Square-root} &
\multicolumn{2}{c}{Logarithmic} &
\multicolumn{2}{c}{Exponential} &
\multicolumn{3}{c}{Sing 3-parameter} &
\multicolumn{4}{c}{Claret 4-parameter} \\
\colhead{} &
\colhead{(K)} &
\colhead{$a$} &
\colhead{$b$} &
\colhead{$c$} &
\colhead{$d$} &
\colhead{$f$} &
\colhead{$g$} &
\colhead{$h$} &
\colhead{$j$} &
\colhead{$k$} &
\colhead{$l$} &
\colhead{$m$} &
\colhead{$n$} &
\colhead{$p$} &
\colhead{$q$} &
\colhead{$r$} &
\colhead{$s$}}
\startdata
5.00 &  4000 &  0.3403 & $-$0.0027 &  0.4552 & $-$0.3431 &  0.9693 &  0.5362 &  0.4500 &  0.2872 & $-$0.0049 &  3.2501 & $-$4.8862 &  2.1927 &  1.3104 & $-$1.0677 &  0.4984 & $-$0.0734 \\
5.00 &  4250 &  0.3448 & $-$0.0012 &  0.4590 & $-$0.3500 &  0.9854 &  0.5428 &  0.4550 &  0.2897 & $-$0.0050 &  3.3255 & $-$5.0239 &  2.2634 &  1.4599 & $-$1.4850 &  0.9751 & $-$0.2612 \\
5.00 &  4500 &  0.3605 &  0.0154 &  0.4579 & $-$0.3180 &  0.9624 &  0.5567 &  0.4502 &  0.3097 & $-$0.0047 &  3.2028 & $-$4.7381 &  2.1097 &  1.1232 & $-$0.4982 & $-$0.1228 &  0.1674 \\
5.00 &  4750 &  0.3930 &  0.0518 &  0.4527 & $-$0.2603 &  0.9267 &  0.5852 &  0.4395 &  0.3476 & $-$0.0043 &  3.0166 & $-$4.2955 &  1.8751 &  0.7786 &  0.4510 & $-$1.0960 &  0.5286 \\
5.00 &  5000 &  0.4355 &  0.1021 &  0.4423 & $-$0.1792 &  0.8719 &  0.6206 &  0.4212 &  0.3973 & $-$0.0038 &  2.7132 & $-$3.6011 &  1.5091 &  0.3962 &  1.4076 & $-$1.9729 &  0.8239 \\
5.00 &  5250 &  0.4690 &  0.1462 &  0.4284 & $-$0.1013 &  0.8090 &  0.6452 &  0.3987 &  0.4379 & $-$0.0032 &  2.3418 & $-$2.7925 &  1.0849 &  0.0812 &  2.0744 & $-$2.4591 &  0.9446 \\
5.00 &  5500 &  0.4908 &  0.1825 &  0.4090 & $-$0.0370 &  0.7486 &  0.6566 &  0.3735 &  0.4647 & $-$0.0028 &  2.0158 & $-$2.1176 &  0.7385 & $-$0.0314 &  2.1191 & $-$2.2464 &  0.7927 \\
5.00 &  6000 &  0.5118 &  0.2485 &  0.3493 &  0.0767 &  0.6173 &  0.6510 &  0.3100 &  0.4932 & $-$0.0022 &  1.5187 & $-$1.1920 &  0.2966 &  0.0026 &  1.5101 & $-$1.1814 &  0.2922 \\
\enddata
\label{tab:R}
\end{deluxetable}

\clearpage
\begin{deluxetable}{cc|r|rr|rr|rr|rr|rrr|rrrr}
\rotate
\tabletypesize{\tiny}
\tablecolumns{18}
\tablewidth{0pt}
\tablecaption{Limb-darkening Coefficients for the Johnson--Kron--Cousins $I$-band}
\tablehead{
\colhead{\logg} &
\colhead{\Te} &
\colhead{Linear} &
\multicolumn{2}{c}{Quadratic} &
\multicolumn{2}{c}{Square-root} &
\multicolumn{2}{c}{Logarithmic} &
\multicolumn{2}{c}{Exponential} &
\multicolumn{3}{c}{Sing 3-parameter} &
\multicolumn{4}{c}{Claret 4-parameter} \\
\colhead{} &
\colhead{(K)} &
\colhead{$a$} &
\colhead{$b$} &
\colhead{$c$} &
\colhead{$d$} &
\colhead{$f$} &
\colhead{$g$} &
\colhead{$h$} &
\colhead{$j$} &
\colhead{$k$} &
\colhead{$l$} &
\colhead{$m$} &
\colhead{$n$} &
\colhead{$p$} &
\colhead{$q$} &
\colhead{$r$} &
\colhead{$s$}}
\startdata
5.00 &  4000 &  0.2865 & $-$0.0260 &  0.4147 & $-$0.3561 &  0.9115 &  0.4669 &  0.4155 &  0.2326 & $-$0.0049 &  3.1293 & $-$4.8514 &  2.2143 &  1.6257 & $-$2.2273 &  1.8286 & $-$0.5969 \\
5.00 &  4250 &  0.2907 & $-$0.0228 &  0.4159 & $-$0.3533 &  0.9134 &  0.4715 &  0.4164 &  0.2367 & $-$0.0049 &  3.1367 & $-$4.8553 &  2.2153 &  1.6239 & $-$2.2141 &  1.8175 & $-$0.5929 \\
5.00 &  4500 &  0.3042 & $-$0.0072 &  0.4131 & $-$0.3167 &  0.8807 &  0.4821 &  0.4090 &  0.2559 & $-$0.0044 &  2.9665 & $-$4.4792 &  2.0142 &  1.1674 & $-$0.8803 &  0.3178 & $-$0.0045 \\
5.00 &  4750 &  0.3321 &  0.0224 &  0.4109 & $-$0.2673 &  0.8502 &  0.5072 &  0.4014 &  0.2890 & $-$0.0040 &  2.7905 & $-$4.0597 &  1.7891 &  0.8079 &  0.1285 & $-$0.7401 &  0.3921 \\
5.00 &  5000 &  0.3687 &  0.0647 &  0.4034 & $-$0.1964 &  0.8016 &  0.5380 &  0.3862 &  0.3324 & $-$0.0035 &  2.5041 & $-$3.4053 &  1.4420 &  0.4247 &  1.1048 & $-$1.6604 &  0.7076 \\
5.00 &  5250 &  0.3985 &  0.1054 &  0.3888 & $-$0.1221 &  0.7384 &  0.5586 &  0.3635 &  0.3692 & $-$0.0030 &  2.1365 & $-$2.6145 &  1.0280 &  0.1100 &  1.7741 & $-$2.1626 &  0.8378 \\
5.00 &  5500 &  0.4175 &  0.1398 &  0.3684 & $-$0.0603 &  0.6777 &  0.5670 &  0.3377 &  0.3930 & $-$0.0026 &  1.8208 & $-$1.9695 &  0.6989 &  0.0098 &  1.7885 & $-$1.9293 &  0.6820 \\
5.00 &  6000 &  0.4340 &  0.1997 &  0.3108 &  0.0434 &  0.5539 &  0.5581 &  0.2774 &  0.4163 & $-$0.0020 &  1.3705 & $-$1.1454 &  0.3099 &  0.0519 &  1.1995 & $-$0.9321 &  0.2201 \\
\enddata
\label{tab:I}
\end{deluxetable}

\clearpage
\begin{deluxetable}{ccc|rrrrrrrrrrrrrrrr}
\rotate
\tabletypesize{\tiny}
\tablecolumns{19}
\tablewidth{0pt}
\tablecaption{Intensity Profiles $I(\mu)/I(1)$}
\tablehead{
\colhead{Filter} &
\colhead{\logg} &
\colhead{\Te} &
\multicolumn{16}{c}{$\mu$} \\
\colhead{} &
\colhead{} &
\colhead{(K)} &
\colhead{0.0000} &
\colhead{0.0667} &
\colhead{0.1333} &
\colhead{0.2000} &
\colhead{0.2667} &
\colhead{0.3333} &
\colhead{0.4000} &
\colhead{0.4667} &
\colhead{0.5333} &
\colhead{0.6000} &
\colhead{0.6667} &
\colhead{0.7333} &
\colhead{0.8000} &
\colhead{0.8667} &
\colhead{0.9333} &
\colhead{1.0000}}
\startdata
$u$ & 5.00 &  4000 &  0.1269 &  0.3821 &  0.4743 &  0.5483 &  0.6097 &  0.6621 &  0.7092 &  0.7517 &  0.7900 &  0.8260 &  0.8594 &  0.8904 &  0.9200 &  0.9481 &  0.9745 &  1.0000 \\
$u$ & 5.00 &  4250 &  0.1384 &  0.3786 &  0.4726 &  0.5467 &  0.6080 &  0.6602 &  0.7073 &  0.7498 &  0.7882 &  0.8243 &  0.8579 &  0.8891 &  0.9190 &  0.9474 &  0.9741 &  1.0000 \\
$u$ & 5.00 &  4500 &  0.1939 &  0.3597 &  0.4507 &  0.5246 &  0.5868 &  0.6404 &  0.6891 &  0.7335 &  0.7738 &  0.8119 &  0.8476 &  0.8808 &  0.9128 &  0.9433 &  0.9720 &  1.0000 \\
$u$ & 5.00 &  4750 &  0.2277 &  0.3261 &  0.4153 &  0.4899 &  0.5540 &  0.6100 &  0.6614 &  0.7086 &  0.7519 &  0.7931 &  0.8319 &  0.8682 &  0.9034 &  0.9370 &  0.9689 &  1.0000 \\
$u$ & 5.00 &  5000 &  0.2301 &  0.2898 &  0.3781 &  0.4545 &  0.5211 &  0.5800 &  0.6345 &  0.6849 &  0.7313 &  0.7756 &  0.8176 &  0.8569 &  0.8950 &  0.9315 &  0.9661 &  1.0000 \\
$u$ & 5.00 &  5250 &  0.2411 &  0.2626 &  0.3521 &  0.4315 &  0.5014 &  0.5632 &  0.6205 &  0.6733 &  0.7218 &  0.7680 &  0.8116 &  0.8524 &  0.8919 &  0.9296 &  0.9652 &  1.0000 \\
$u$ & 5.00 &  5500 &  0.2190 &  0.2446 &  0.3349 &  0.4172 &  0.4899 &  0.5542 &  0.6135 &  0.6680 &  0.7179 &  0.7652 &  0.8097 &  0.8511 &  0.8911 &  0.9292 &  0.9651 &  1.0000 \\
$u$ & 5.00 &  6000 &  0.1891 &  0.2327 &  0.3175 &  0.4001 &  0.4752 &  0.5422 &  0.6042 &  0.6610 &  0.7128 &  0.7617 &  0.8074 &  0.8498 &  0.8905 &  0.9290 &  0.9650 &  1.0000 \\
\enddata
\end{deluxetable}

\end{document}